# Polarized Beams in a Muon Collider

B. Norum[a] and R. Rossmanith[b]

[a]University of Virginia, Charlottesville, USA

[b]DESY, Hamburg, Germany

This paper presents a first overview on how to obtain polarized beams in a high energy muon collider. Depolarization due to cooling as a function of the cooling energy is estimated. The spin handling system during acceleration and final storage is sketched out and a first idea on the possible polarimetry is described.

## 1. INTRODUCTION

In various recent publications new concepts for a high energy muon collider was presented [1,2]. It is claimed that a muon collider could be ideal for high energy, high luminosity lepton collisions. In this paper the possibility of obtaining polarized muon beams is discussed.

Muons are generated during a pion decay. The pion decays via the weak interaction into a muon and a muon neutrino (fig. 1). The muons are born polarized. In order to obtain polarized collisions the polarization has to be maintained during acceleration and storage. This means that the polarized muons have to be extracted from the decaying muons, pre-accelerated and cooled with little loss of polarization and polarization has to be maintained during acceleration and in the collider. The polarization direction in the interaction region has to be longitudinal and, in order to avoid systematic errors, the direction of polarization has to reversed from time to time. Polarization has to be measured at various positions in the accelerator.

It is obvious that it will be impossible to cover the above mentioned-topics in great detail in this paper as the layout of the whole accelerator is not yet firmly established. All the topics mentioned in this paper need more detailed simulations. The aim of this paper is to initialize a discussion with regard to the feasibility of maintaining polarization and to list the topics requiring more detailed investigations.

Polarized muons are not a new subject in physics. Muon storage rings for the measurement of the anomalous magnetic moment of the muon exist at various institutions [3]. Polarized muons are used as a tool to investigate materials [4]: the spins of the muons are rotated by the internal fields of the probe. And finally the 'spin-crisis' effect was first measured by the SMC group using polarized muons [5].

Nevertheless, there is a fundamental difference between these applications of muons and the use of muons in a high energy muon collider: the high luminosity requires high current and low emittance beams which have never produced before. All the proposed manipulations on obtaining these beams must be designed in such a way that they do not effect polarization.

## 2. THE FUNDAMENTAL CONSTANTS

The rest mass of the muon is 105.66 MeV and therefore about 207 times larger than the rest mass of the electron. The anomalous magnetic moment of the muon is

$$\left(\frac{g-2}{2}\right)_\mu = 1.166 \times 10^{-3}$$

compared to the electron anomalous magnetic moment of $1.16 \times 10^{-3}$ [6].

Since the lifetime of the muon is relatively short ($2.2 \times 10^{-6}$ sec), a selfpolarization mechanism at high energies is almost excluded: the muons have to be collected polarized and polarization has to be maintained during acceleration and storage.

During acceleration the integer resonances occur when (n = integer)

$$\left(\frac{g-2}{2}\right)\gamma = n$$

This formula for n=1 gives the distance between two integer resonances. The distance between two integer resonances is 94.62 GeV compared to ca. 440 MeV with electrons and ca. 470 MeV with protons.

The figures clearly show that the muon spin is extremely insensitive to magnetic fields.

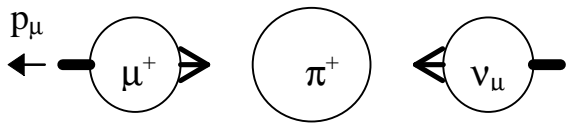

Figure 1. The generation of a muon from a pion decay. The muon is polarized opposite to its direction of motion.

This has a positive aspect. Polarization is far more stable than it is in the case of electrons and protons. However, there is also a negative aspect: in the relativistic limit an integrated field of 492 Tesla.m is required to rotate the muon spin by 90 degrees. In comparison, 2.3 T.m are required to rotate the spin of an electron by 90 degrees and 2.7 Tm for rotating the spin of a proton by 90 degrees.
The clear advantage is that even at an end energy of 2 TeV the spin tune is only 21.13 and comparable to a ca. 10 GeV electron storage ring [7].

## 3. THE MODEL MUON COLLIDER

As mentioned in the introduction, the layout of the muon collider is not yet completely established. As a basis for the following considerations, fig2 [8] represents a possible model fo the design of the machine. A proton beam hits a target and produces pions. The pions decay into muons, the muons are pre-accelerated and transported to the cooling section. The cooling section consists of a block of Be or Li and an energy restoring cavity. The damping mechanism is similar to the radiation damping in an electron storage ring. The only difference is that the muon energy is reduced by colliding with an electron. As a consequence, the degree of polarization of the muon beam can be reduced.

After damping the whole beam is accelerated to its maximum energy either by a circular type of accelerator or a recirculator. Due to the short lifetime the time, in which the muon is accelerated, defines the luminosity.

After acceleration the muons are stored in a storage type device and brought to collision. In the interaction region, the spins have to aim in the longitudinal direction. Two versions of the muon collider are discussed: one with an end energy of 300 GeV and one with an end energy of 2 TeV.

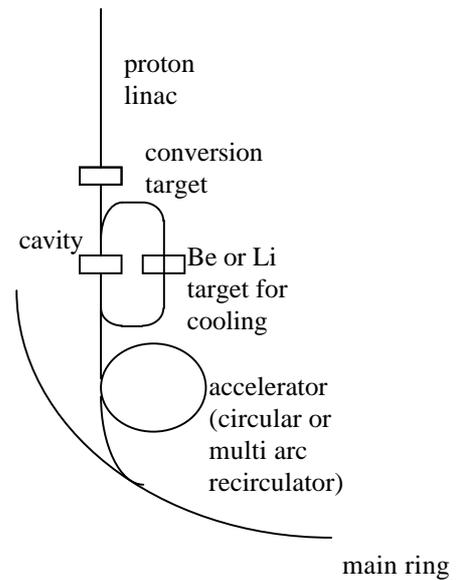

Figure 2. Schematic layout of the muon collider with conversion target, cooling section, accelerator and collider ring

The main reasons for depolarization are:
a.) Depolarization in the decay channel. According to fig. 1 spin and momentum aim in the same direction. In order to obtain a beam with a high degree of polarization, particles with a certain momentum have to be selected. This effect reduces the maximum polarized current and therefore the polarized luminosity.
b.) Cooling can be very deleterious for polarization. Due to muon-electron collisions the muon beam can be depolarized.

c.) During acceleration to 300 GeV or 2 TeV the beam has to cross several depolarizing resonances (3 integer resonances for an end energy of 300 GeV, 21 resonances for an end energy of 2 TeV). Depolarization has to be minimized. One serious problem is that the classical remedies against depolarizing resonances used at electrons and protons like Siberian Snakes [9] have only limited application in a muon collider. As mentioned in chapter 2 the required integrated fieldstrength for such a device is very high.

d.) Depolarization during storage has to be minimized.

## 4. THE GENERATION OF POLARIZED MUONS

The pion rest mass is 139.6 MeV [6]. When the decaying pion is at rest the kinetic energy of the muon is 4.1 MeV according to the laws of energy and momentum conservation.

In the muon collider the muons are produced in a so-called decay channel by moving pions. When the pion moves the momenta of the muon and the pion add or subtract ('forward' or 'backward' muon) [10].

$$p_\mu = \frac{(m_\pi^2 - m_\mu^2)(p_\pi^2 + m_\pi^2 c^2)^{1/2} \pm p_\pi (m_\pi^2 + m_\mu^2)}{2 m_\pi^2}$$

Forward and backward muons have different spin directions. The plus and the minus sign describes the two possibilities.

The formula will be explained in the following by an example. When it is assumed that the pions are monochromatic and have a momentum of 200 MeV/c, muons with a momentum of 209 MeV/c (forward muon) or 105 MeV/c (backward muon) are produced. The muon beam is unpolarized. Polarization is obtained by energy selection.

The intensity of the polarized muon beams generated in a decay channel is described by the magnitude: number of polarized muons per $cm^2$ and sec (luminosity). The luminosities thereby achieved are in the range of $10^5$ to $10^7$. The degree of polarization is close to 100% [11].

The muon collider will operate with $2 \times 10^{12}$ muons per bunch and a repetition rate of 30 Hz in order to obtain a luminosity of $10^{35}$ at 2 TeV. The overall yield is 0.22 muons per initial proton and the muons are produced in a decay channel with a solenoid field. In order to obtain this high intensity, muons within a high energy range have to be captured. The energy of the produced muons lies between 0 to 3 GeV and can only be captured by a special linac which unifies the energies of the muons. From previous statements it is clear that such a beam is unpolarized. The momentum of the forward and backward muons as a function of the momentum of the pion is shown in fig. 3.

In order to obtain polarized muons, the high energy muons have to be selected (fig. 3). This reduces the maximum polarized current to ca. 1/5 of the unpolarized current or the polarized luminosity to ca. 1/25 of the polarized luminosity. The exact number has to be calculated by taking the initial pion distribution and the efficiency of the capturing linac into account.

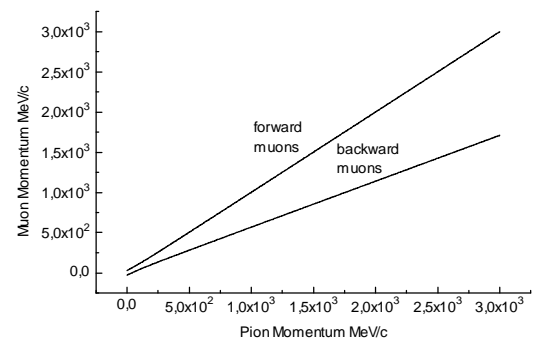

Fig. 3 Momenta of backward and forward muons as a function of pion momentum

The only way to increase the efficiency is to apply a tranverse field along the decay channel and separate the pion trajectories according to their energy. The decay channel is curved and the forward and backward muons can be separated in a more efficient way. Again, a more detailed study is necessary. Nevertheless, even when assuming the best of all scenarios at least 50% of the particles get lost thereby reducing the polarization by a minimum factor of 4. In a more realistic approach the reduction factor of the polarized luminosity will be between 10 to 50 depending on the initial energy distribution of the pions and the layout of the decay channel.

## 4. SYNCHROTRON SPIN MATCHING FOR HORIZONTALLY POLARIZED BEAMS

The second big obstacle for the polarized muon beam is the cooling section (see fig. 2).

After generating the polarized beam it has to be decided whether the spin direction should be horizontal or vertical. The captured muons are brought to a couple of phase-rotating linacs which reduce the energy spread of the muon beam. After leaving these linacs the beam has an energy of between 70 and 400 MeV (depending on the particular cooling scenario) and an energy spread of about 15% [8].

In the case of longitudinal polarization the spins will differ by

$$\left(\frac{g-2}{2}\right)\gamma \cdot 360 \text{ degrees}$$

after one revolution. After 100 revolutions in any of one of the accelerators, the horizontal muons will have suffered severe depolarization.

At low energies it seems to be difficult to rotate the spin into the vertical direction. It seems to be easier to keep the polarization in the horizontal plane as long as possible and to rotate it at higher energies when the bending fields become higher.

The maximum energy in the cooling section is 400 MeV, the energy spread therefore 60 MeV (15%). In the following it is assumed that many passes through the cooling element are needed to damp the emittance.

Ignoring for a moment the energy-loss in the cooling material, the overall depolarization is zero when the beam stays in the accelerator for one synchrotron period (or for n-periods). In the following this technique is called synchrotron spin matching.

In the case of synchrotron spin matching the number of revolutions, the momentum compaction factor and the acceleration voltage all have to be matched appropriately. The matching condition is

$$\int \left(\frac{g-2}{2}\right)\Delta\gamma(t)dt = 0$$

in the relativistic limit where $\Delta\gamma$ is the energy deviation. It is obvious that this condition is fulfilled after a synchrotron oscillation period where

$$\Delta\gamma = \gamma_0 \sin\omega_s t$$

and

$$\int_0^{2\pi} \Delta\gamma(t) = 0$$

In a more general sense the synchrotron spin matching condition can also be fulfilled for a non-periodic function when

$$\int_0^{t_1} \Delta\gamma(t) = 0$$

In order to minimize depolarization a particle with 15% energy deviation (and an energy of 400 MeV) has to gain 120/2.n MeV when it is ejected after 2n revolutions. Momentum compaction, cavity voltage and RF frequency have to be designed in such a way that this condition can be fulfilled.

Synchrotron spin matching becomes more difficult during cooling. In the course of this process the particle energy is reduced by an unpredictable amount in the cooling device. Since the distance between the cooling device and the cavity is known the effect can be compensated. This is not possible for synchrotron radiation cooling since the position where the particle looses energy is not known.

It is obvious that a compensation scheme shown in the following is not required when the number of revolutions is small. Nevertheless, an example is given to demonstrate the efficiency of synchrotron spin matching. In fig. 4 the particles pass, before they enter the cooling element, a section with a negative momentum compaction factor $\alpha$ and afterwards a section with a positive $\alpha$. The absolute amount is the same. As a result the length of the trajectory depends only on the amount of the lost energy in the cooling element. When $\alpha$ has the correct value synchrotron spin matching is possible even with cooling.

After cooling, in the accelerator, synchrotron spin matching has to be applied in order to minimize depolarization.

## 5. DEPOLARIZATION IN THE COOLING ELEMENT

In order to estimate the degree of depolarization during ionization cooling, it is assumed that the binding energy of the electrons is small compared to the muon energy: the electrons are considered as free and unpolarized particles.

The spin flip probability during scattering of polarized Dirac particles on electrons was first calculated by Ford and Mullin [12]. In the nonrelativistic approach, the probability Q that a muon is scattered into the angle θ is given by

$$Q(\varepsilon,\theta) = \frac{1+\varepsilon}{2} - \varepsilon \frac{m_e^2}{m_\mu^2} \beta^4 \cdot [\sin^2(\theta/2) - \sin^4(\theta/2) - \sin^6(\theta/2)]$$

θ is the center of momentum angle. ε is +1 when the spin remains the same after scattering. ε is -1 when the spin is flipped. $m_e$ is the electron rest mass and $m_\mu$ is the rest mass of the muon. β is v/c in the usual definition.

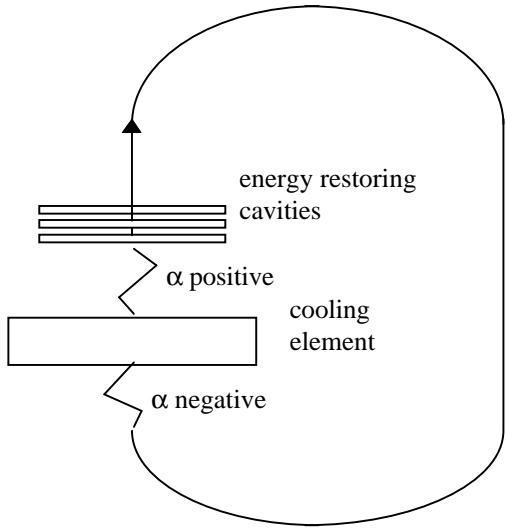

Fig. 4 Sections with positive and negative momentum compaction factors in front and after the cooling element compensate the different spin motions for particles with different energy losses in the cooling element

The magnitude

$$w = \left(\frac{m_e}{m_u}\right) \beta^2 \sin^2(\theta/2)$$

is the fractional energy loss of the muon in the lab frame. Comparing both equations it can be observed that the spin flip probability is proportional to the fractional energy loss. The proportionality constant is

$$\left(\frac{m_e}{m_u}\right)\beta^2$$

It becomes clear that depolarization is almost negligible when β is small. In the relativistic approach a similar equation holds. The equations have to be multiplied by a function F(γ). For the practical purposes considered in this paper this is very close to 1 and hence can be neglected.

The muon polarization is very stable compared to electron polarization when in both cases the scattering partner is an electron. This fact is very well known in the muon spin rotation technique. Taking the available simulation datas on ionization cooling from the paper of Van Ginneken, a 500 MeV muon beam looses 3.5 MeV in a 1 cm thick Be target [11]. The corresponding depolarization is ca. $5 \cdot 10^{-5}$ according to the above-introduced formulae.

More generally, the depolarization can be described by the following formula

$$P = P_0 \exp[-x/a]$$

P is polarization, x the length of the particle trajectory in the material. a is the decay length of the polarization. For Be and the given energy a is about 200 m. The simple exponential formula is not completely true since the particle changes the energy and a depends on β. Nevertheless, for all practical purposes the depolarization during cooling is very small and can be conserved when the cooling is carefully executed.

One would expect that Coulomb scattering greatly contributes to depolarization. The cross section for Coulomb scattering is almost Z times larger than the electron cross section. Rose and Bethe [13] have shown that this is not true: this effect is almost negligible.

Summarizing this chapter: depolarization during cooling can be a small effect when the parameters are selected carefully. These findings are in good agreement with the experimental results obtained in the muon spin rotation experiments.

## 6. ACCELERATION OF COOLED POLARIZED MUONS

The polarization of the cooled muons is in the horizontal plane. As mentioned before, a field of 492 Tesla.m is required to rotate the spin from the horizontal direction into the vertical.

Two effects can cause depolarization during acceleration:

a.) the energy-spread leads to depolarization when the spin is still in the horizontal plane b.) during acceleration several resonances have to be crossed and the beam becomes depolarized.

Due to the short lifetime of the muons a CEBAF type recirculator was proposed to accelerate the muons. The linac accelerates the muons by 100 GeV. In such a recirculator depolarization due to spin resonances is fairly small. The arcs are only crossed once at a certain energy and when the harmonic content of the trajectory is small then the effect on polarization is small. Depolarization in this case is caused by the interaction between the deflecting fields of the bending magnets and the quadrupoles in between as shown in fig. 5.

A vertical betatron oscillation acts on the spins proportional to the integral

$$\int_0^\pi B_{quad}(s) \frac{\sin\varphi_{spin}}{\cos\varphi_{spin}} ds$$

where $B_{quad}(s)$ is the horizontal magnetic field of the quadruoles, s is the coordinate along the trajectory and $\varphi_{spin}$ is the phase advance of the spin in the bending magnets along the trajectory. The integral is a simple Fourier integral and can be minimized when the optics is chosen accordingly.

Depolarization due to this effect is very small especially when the emittances are becoming small due to adiabatic damping.

The statements are true for both the vertical and horizontal spin direction.

The main depolarization stems from the fact that the spins are still in the horizontal plane. In order to overcome this depolarization effect the technique described in fig. 4 (synchrotron spin matching) can be applied. Particles which have an energy deviation in one of the arcs will have an opposite energy deviation after the next acceleration when the momentum compaction factor in the arc is chosen accordingly. The correct choice of arc length and momentum compaction factor allow to maintain polarization during acceleration.

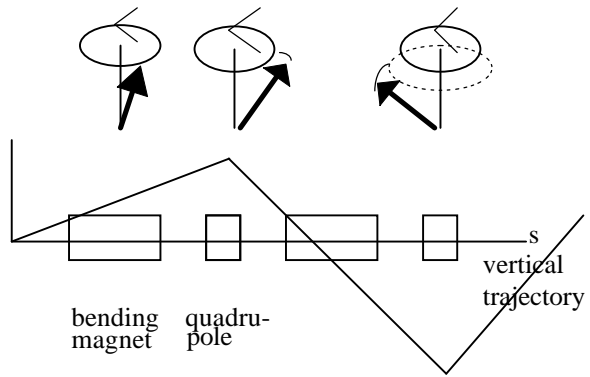

Fig. 5 The action of a vertical betatron oscillation on the spin direction. The sequence of spin rotations in the bending magnets around the vertical axis and the spin rotations in the quadrupoles around the horizontal axis can change the spin direction. In the case where the trajectories are different for different particles, the beam becomes depolarized. The effect can also be used to rotate the spin.

Summarizing these chapters: polarization is not affected during acceleration even when the spins remain in the horizontal plane during the whole acceleration process. A prerequisite is that the arcs are designed to avoid depolarization. This is done by selecting the correct momentum compaction factor in each arc.

# 7. SPIN HANDLING IN THE STORAGE RING

Two topics have to be considered when discussing polarization in the storage ring.
a.) Spin rotation into the longitudinal direction around the interaction region.
b.) Rotation of the spin of the injected muons into the vertical direction.

A general type of spin rotator for muons is shown in fig. 6. It is assumed that the field strength of the bending magnets will be ca. 10 T and that the magnets are about 10 m long . As a result, a 30 m long magnet assembly consisting of 3 magnets is able to rotate the spin by 45 degrees.

In the arc additional bending magnets deflecting the beam vertically (3 at the beginning and 3 at the end of the spin rotator) are installed. Between these additional bending magnets a 120 m long section of the normal bending magnets of the storage ring

This design which can be modified in various ways has the added advantage that it shields the interaction region from the arcs. In this design the additional required length is only 60 m per rotator (10 Tesla magnets assumed). The well known spin matching conditions have to be applied between the two spin rotators (one before and one after the interaction region).

The spin can be rotated from the horizontal into the vertical direction just before it enters the main ring as described in fig. 6.

Since there is no acceleration in the main ring, depolarization by the effects shown in fig. 5 will remain fairly modest. The only source of depolarization could be the region where polarization is horizontal. The spin matching condition must be valid for all particles independent of its energy and its position in the transverse phase space. As a result, the design of the interaction region may require special attention. Since the spin rotators have to be part of the machine they have to be taken into account from the beginning.

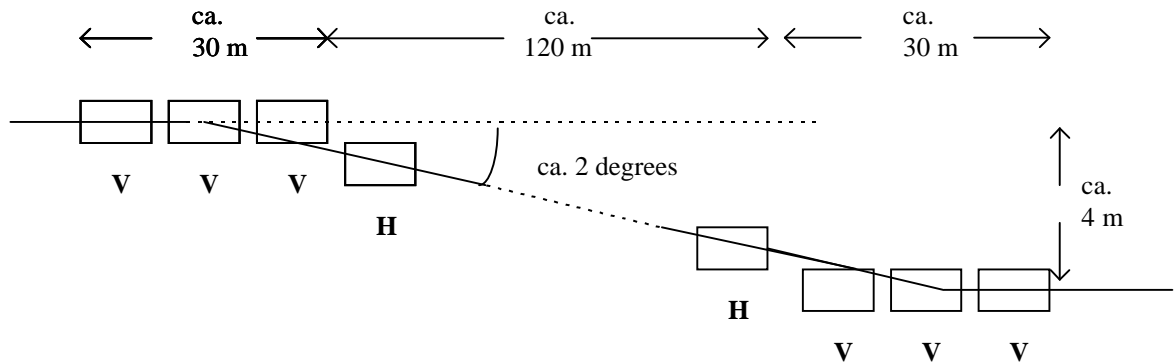

Fig. 6 Possible spin rotator for muons in the main ring. The spins are rotated 45 degrees from the vertical towards the momentum axis by the first 3 ca. 10 T, 10 m long vertically deflecting magnets. The spin is afterwards rotated by 180 degrees around the vertical axis by 12 normal bending magnets and finally into the longitudinal direction by the last three vertically deflectling magnets. The additional space requirement for the spin rotators is 120 m on each side of the interaction region. H are horizontally deflecting magnets and V vertically deflecting magnets.

(spin rotation around the vertical direction) is tilted by ca. 2 degrees followed by a 30 m long vertical bend into the opposite direction.

# 8. THE CHANGE OF HELICITY IN THE INTERACTION REGION

Up to now only particles of one sort were discussed. In fig. 1 the decay of a $\pi^+$ was shown producing a $\mu^+$. $\mu^-$ are produced by the decay of $\pi^-$ mesons. Negative muons have positive helicity since they are born from a negative muon together with an anti-neutrino.

The first difference shows during cooling. Positive muons do not interact with the nucleus, negative muons are attracted by the nucleus and can form muonic atoms at low energies when they are significantly slowed down. This property is mainly used in solid state physics where the angular momentum of the muons can be measured when they decay.

When the beam energy remains much higher than the binding energy the muons can be considered as free and the difference between positive and negative muons is small. This is especially true when the cooling is performed in a material with a low Z.

There is no difference between positive and negative muons during acceleration: the momentum compaction factors in the arcs are the same when they are passed clockwise or counter-clockwise.

Fig.7 shows a possible scenario for the arrangements of the spin rotators in the ring. The first spin rotator rotates the spin from the vertical (upward) into the longitudinal direction . After passing the interaction region and the second spin rotator, the spin aims in the vertical (downward) direction. When the particle returns to the interaction region after one revolution its spin aims into the opposite direction compared to the first pass: the spin direction changes from revolution to revolution.

A similar statement is true for the opposite particle. As a result the helicities of the colliding bunches can be altered from revolution to revolution.

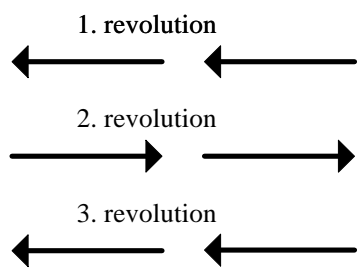

The injected beam has to be added to the beam in such a way that the spins of the muons add up. In other words the injection has to be performed every second revolution.

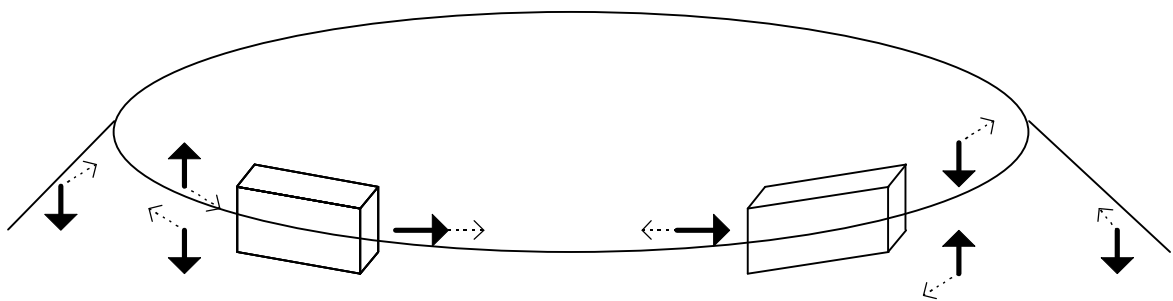

Fig. 7 shows a possible scenario for arranging the spin rotators in order to obtain varying helicity directions from interaction to interaction. After the particles have passed the two spin rotators surrounding the interaction region, the spin aims in the opposite direction and changes its direction the next time it passes the interaction region. This means that polarized interactions with a low systematic error with a low systematic error can be obtained.

By selecting the correct injection time, the relative spin direction of the two bunches can be altered and the systematic error can be reduced. This is especially an attractive solution as the lifetime of the muons is limited and changes can be made within seconds simply by changing the timing of the injection.

The spin rotation between the multi-arc accelerator and the main ring is performed by a spin rotator similar to the one used in the ring and can be integrated into the connecting arcs between accelerator and storage ring.

## 9. POLARIMETRY

The muons decay into positrons or electrons

$$\mu^+ \rightarrow e^+ \nu_e \bar{\nu}_\mu$$

$$\mu^- \rightarrow e^+ \nu_\mu \bar{\nu}_e$$

The distribution of the electrons and the positrons is commonly used as a polarimeter reaction for muons. In the main storage ring the situation is different compared to the normal situation in solid state physics. Most of the electrons or positrons rapidly loose energy due to their synchrotron radiation and will hit the inner vauum chamber in the bending magnets.

The lifetime for a non-relativistic muon is 2.2 μsec, γ is ca. $10^4$ and therefore the life time is 0.22 sec. One bunch contains $10^{12}$ particles. Per meter ca. $1.7 \times 10^4$ muons decay per second.

In the rest system of the $\mu^+$ the positron is mainly emitted in the forward direction as a consequence of parity violation. The decay probability is

$$W(\theta) = 1 + a_0 \cos\theta$$

where θ is the angle between the spin direction and the positron trajectory. $a_0$ depends on the energy and is 1/3 when all positron energies are taken into account. In the ring the spins aim in the vertical direction. The positrons are mainly emitted into the spin direction. In the lab frame the preferred emission angle θ becomes therefore 1/γ or $10^{-4}$ rad. This is a reasonable angle for a polarimeter.

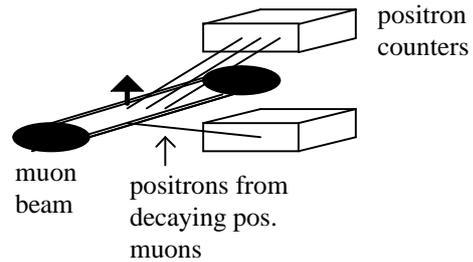

Fig. 8 A possible layout for a high energy muon polarimeter in the main ring. The asymmetry of the positrons is measured.

The polarimeter consists of a positron detector surrounding the vacuum chamber. The asymmetry is very high compared to Compton polarimeters. In a realistic approach it could be as high as 2 to 1 (depending on the energy selection of the particles and the acceptance of the polarimeter). It should be possible to measure polarization in a few seconds with an accuracy of 1%.

A similar statement is true for negative muons. Instead of positrons, electrons are measured. Fig. 8 shows a possible layout of the polarimeter in the main ring. Additional magnets (not shown in the figure) can improve the separation between muon beam and positrons (electrons).

At lower energies the spin is in the horizontal direction. The distribution W(θ) makes it somewhat difficult to measure the spin aiming in the momentum direction. Fortunately this is not necessary since the spins rotate around the vertical axis in the arcs. The polarimeters should be placed at positions where the spin is not longitudinal but radial. The polarimeter shown in fig. 8 has to be rotated by 90 degrees.

## 10. SUMMARY

The muon collider is in ideal tool to study polarized lepton collisions at high energies. For this purpose the muon collider is much better than an electron collider. The polarization is much more stable due to the higher mass of the muon and an almost identical (g-2)/2. The high mobility of the spins of high energy electrons and also of protons requires enormous accuracy in the adjustments of the orbits in order to handle the spins. Muon polarization is

robust with regard to orbit distortions and misalignments. There are 2 drawbacks however:

a.) The most severe is that the selection of polarized muons in the decay channel limits the maximum polarized luminosity. The decay channel is the bottle neck of the whole machine and needs more detailed investigation.

b.) The spin handling system poses difficulties, especially in the low energy region. This is the other side of the coin of the high stability of the polarization.

Nevertheless, if the interaction region is intelligently designed, double spin experiments with a low systematic error are possible. This would offer an enormous advantage over electron-positron linear accelerators where double spin experiments are almost completely out of the question.

It is interesting to see that ionization cooling does not really harm polarization when the beam to be cooled has a low enough energy so that the fractional energy loss is not too high. On the other side, the energy has to be high enough so that negative muons cannot be captured by the positive charged nuclei of the cooling material. In addition this effect can be minimized by using a cooling material with a low Z.

The use of multiple arcs eliminates the danger of resonance depolarization during acceleration: the arcs are only transversed once. Nevertheless, the arcs have to be designed carefully due to the fact that the spin is in the horizontal plane. Depolarization is compensated by chosing the momentum compaction factor in an appropriate way. In the next pass through the linac the particles which had less energy gain more energy and the spins become parallel in the next arc. This technique is called synchrotron spin matching. Only at the highest energies are the spins rotated into the vertical direction.

Polarimetry seems to be relatively easy.

## ACKNOWLEDGEMENTS

The authors have to thank many people for their support, especially Prof. D. Cline from UCLA and Prof. B. Palmer for stimulating the work on this paper. One of the authors (RR) has to thank UCLA for financial support and the Forschungszentrum Karlsruhe (and in the Forschungszentrum especially Dr. H. O. Moser) for support and interest.

This paper could not have been written without critical comments and discussions with many colleagues, especially D. Neuffer from FNAL, Y. Yokoya from KEK, A. Chao from SLAC, Y. Derbenev from DESY, J. G. Gallardo from BNL and many others.